\documentclass[11pt,twoside]{article}
\usepackage{asp2010}
\usepackage{epsf}
\usepackage{graphicx}
\resetcounters
\begin{document}
%\bibliographystyle{aj}
%\renewcommand{\thefootnote}{\fnsymbol{footnote}}

%%% FILL IN THE TITLE OF THE ARTICLE
\title{The Limb Brightening and its Relationship with the Millimeter-wave Cavity}
%\title{\footnotemark}

%%% FILL IN AUTHOR AND AFFILIATION INFORMATION
\author{V.~De~la~Luz,$^1$ ~M.~Chavez,$^1$ and ~E.~Bertone$^1$
\affil{$^1$Instituto Nacional de Astrof\'\i sica \'\O ptica y Electr\'\o nica,
Luis Enrique Erro 1, Sta. Ma. Tonantzintla, Puebla, Mexico}}

%\author{
%  \affil{}
%  }

%%% ADD SUMMARY OF ARTICLE
\begin{abstract}%INSERT ABSTRACT TEXT ON THIS LINE (NO CARRIAGE RETURN!)
Through a detailed theoretical analysis of the local emission at
millimeter,sub-millimeter and infrared wavelength regimes (from ∼ 10 GHz up to
∼ 10 THz), we found that, associated with the temperature minimum, there is an
optically thin cavity surrounded by two regions of high local emissivity. We
call this structure the Chromospheric Solar Millimeter Cavity (CSMC). In order
to search for traces of this cavity in the available radio observations on the
solar limb, we have developed a robust method to associate the radiation at
different heights with the observed brightness temperatures. We foresee that
this approach will allow us to determine the relationship between the CSMC and
the solar limb brightening.
\end{abstract}
%\footnotetext{}
%\renewcommand{\thefootnote}{\arabic{footnote}}
%\setcounter{footnote}{0}

%%% INSERT MAIN BODY OF ARTICLE HERE. CONSULT "PREPARING YOUR ARTICLE FOR
%%% THE ASP CONFERENCE SERIES", SECTIONS 3.8, 4 AND 5 FOR HELP WITH EQUATIONS, FIGURES,
%%% AND TABLES.

\section{Introduction}
\label{generalintroduction}
The first observations of the solar radio emission 
shown a high brightness temperature  hundreds
of times grater than the observed thermal radiation from the photosphere
\citep{1947RSPSA.190..357M}. This radio emission was associated with the solar
Corona region \citep{1946Natur.158..632M}. 

In \cite{1950AuSRA...3...34S} the solar limb brightening and the solar radii
below 24 GHz was theoretically predicted. This early work, shows
that the solar radii grows in diameter while the frequency of the emission
decreases.  The observed apparent temperatures was associated with the
chromosphere and coronal regions. 

Several observations at high frequencies were performed in order to
characterize the limb brightening \citep[We can found an extended review about
  the observations at short millimeter range in][]{1981SoPh...69..273A}. These
works show that the theoretical limb brightening could be affected by the
spicules. However, \cite{1981SoPh...69..273A} shows that the height
scale of the transition zone model is unimportant. Is important remark that
these first models present higher temperature minimum than the actual values.
%The temperature minimum used in their model is higher than the actual values. 
The temperature minimum plays a fundamental role in the chromospheric emission
because it is requiered to reproduce observations of CO moleculae. The CO have
a threshold around 4200 K to avoid their desintegration 
\citep{1978ApJ...225..665A}.
%The results also includes a correlation between
%the limb brightening at millimeter wavelengths and the solar cycle.

The relation between the limb brightening and the microstructure (spicules)
was studied in detail in \cite{2005A&A...433..365S}. However, the
approximation of full ionizated chromosphere and the empirical model of the
chromosphere is not a good approximation at millimeter region
\citep{2011ApJ...737....1D} besides that their minimum of temperature is higher
than 4200 K. 

%The limb brightenning model includes a chromospheric and coronal model of
%temperature, density and pressure. The chromosphere is the responsible of
%the observed radii at millimeter, sub-millimeter and infrared wavelengths
%\cite{}. In these sense, the chromospheric model are very important for
%the determination of the solar radii.
%In the literature we found several families
%of solar chromospheric models \citep{1973ApJ...184..605V,
%  1990ApJ...355..700F,1992ASPC...26..499C}. In particular, the C7 model from
%\cite{2008ApJS..175..229A} introduces new physical mechanism to cool the high
%chromosphere (ambipolar diffusion), this region is the responsible of the
%emission at millimeter wavelengths. In \cite{} we show that the
%both C7 and VALC models create high emission at sub-millimeter wavelengths.
%In the case of the millimeter emission the C7 models compute low brightness
%temperatures. The VALC models generate high emission than observed. We also
%found that the clasical bremmstrahlung emission is only important in the high
%chromosphere (500 km over the photosphere) while the H- emission is the most
%important mechanism in the low chromosphere. 

In De la Luz et al 2012 (under review) we found that the temperature minimum
of the chromosphere create a structure called Chromospheric Solar
Millimeter-wave Cavity (CSMC). This structure is the responsible of the
morphology of the solar spectrum at sub-millimeter wavelengths. Inside of this
structure the Bremsstrahlung emission is the most important emission mechanism.

%This work studies the emission in the center of the solar disk. 
In this work we study the relation between the CSMC and the solar radii using
three opacities functions and NLTE computations using the C7 model
\citep{2008ApJS..175..229A} that introduce the minimum of temperature at 4200 K.

We presents a theoretical limb brightening function from 2 GHz to 1 THz
using our model PakalMPI \citep{2010ApJS..188..437D} and show the behavior of
the CSMC close to the limb. 

%We introduce the model (Section \ref{themodel})

\section{The model}\label{themodel}  
We are using PakalMPI to solve the radiative transfer equation in Non-local
Thermodynamical Equilibrium using the C7 model for the chromosphere \cite{2011ApJ...737....1D}. 
We compute three opacity functions to reproduce the spectrum in the continuum: The Wildt's photo-detachment mechanism \citep{1939ApJ....90..611W},
The neutral interaction \citep{1996ASSL..204.....Z}, and The Bremsstrahlung for Hydrogen negative ion \citep{1980ZhTFi..50R1847G}.

PakalMPI computes the brightness temperature for different position in the
solar disk in NLTE using the approximation for the departure coefficient (b1)
defined by \cite{1937ApJ....85..330M} %and modify in
                                %\cite{1973ApJ...184..605V} 
$$
b_1 = \frac{n_1/n_1^*}{n_k/n_k^*},
$$
where the ratio $n_1^*/n_k^*$ is given by the Saha equation in thermodynamical
equilibrium.

We compute the electronic density using \cite{2011ApJ...737....1D}
%\begin{equation}\label{elecde}
$$
n_e= \frac{-(1-Zd)+\sqrt{(1-Zd)^2+4d(n_HZ)}}{2d},
$$
%\end{equation}
where $n_H$ is the hydrogen density, $Z$ is the ionization contribution 
function, $T$ the temperature and $d$ the  contribution in NLTE.
The ionization contribution function is
$$
Z=\sum_{\xi = He}^{\Xi}\sum_{N_{\xi}=1}^{F} N_\xi n_{\xi,N_\xi},
$$
from \citep{1973ApJ...184..605V} where $N_\xi$ is the ionization stage and
$n_{\xi,N_\xi}$ is the ion density. The contribution in NLTE is
\begin{equation}\label{defd2}
d=b_1\psi(T),
\end{equation}
where $\psi(T)$ is the classical Saha equation.

\section{The CSMC at Limb}
\begin{figure}
\begin{center}
\includegraphics[width=1.0\textwidth]{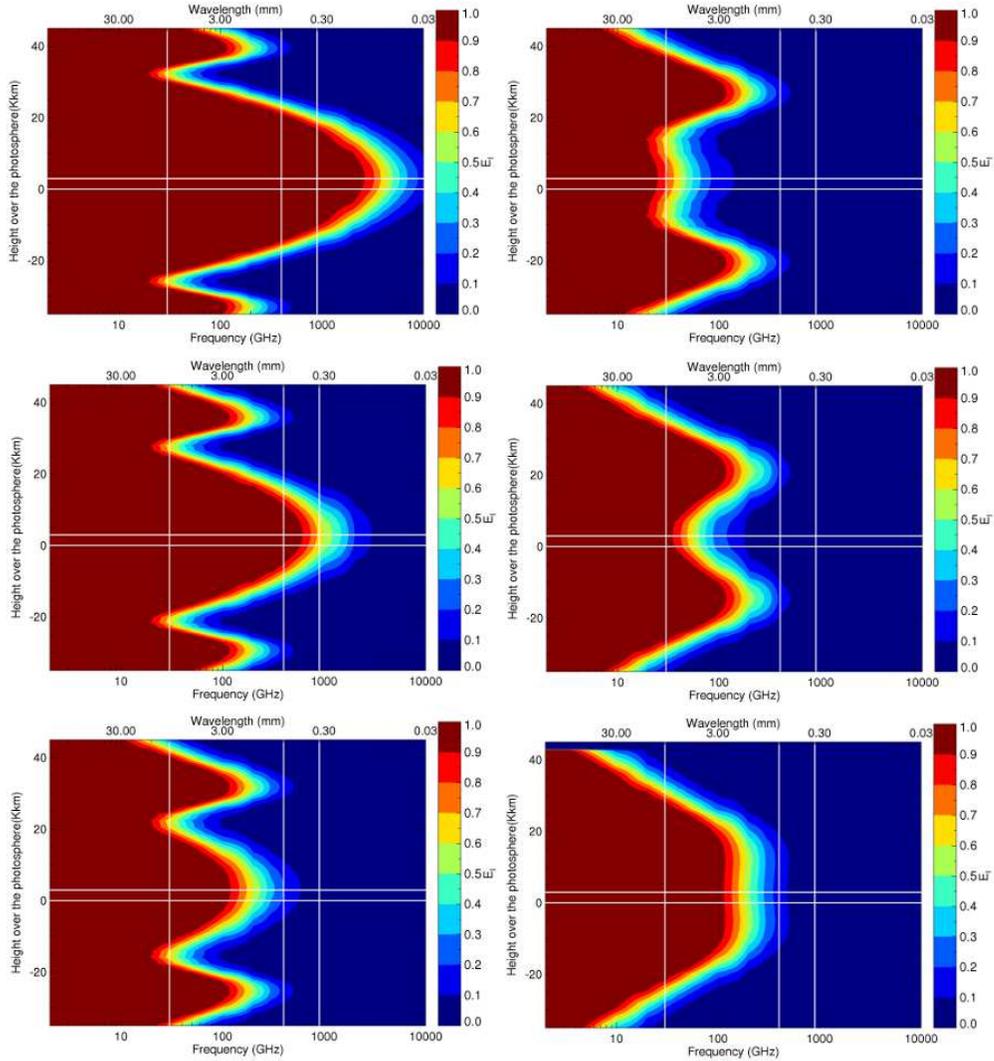}
\caption{Local emission efficiency of the solar limb from 127 km (upper-left)
  to the 1018 km over the photosphere (bottom-right). 
  We found three   optical thick regions, while the computations move to the
  corona, these three regions move to the center and at the same time down in
  frequency. The 6th sub-figure (from up to down and from left to right) shows
  the at 840 km over the photosphere the two regions of emission merges in a
  single region.}\label{figure1.ps}  
\end{center}
\end{figure}

\begin{figure}
\begin{center}
\includegraphics[width=1.0\textwidth]{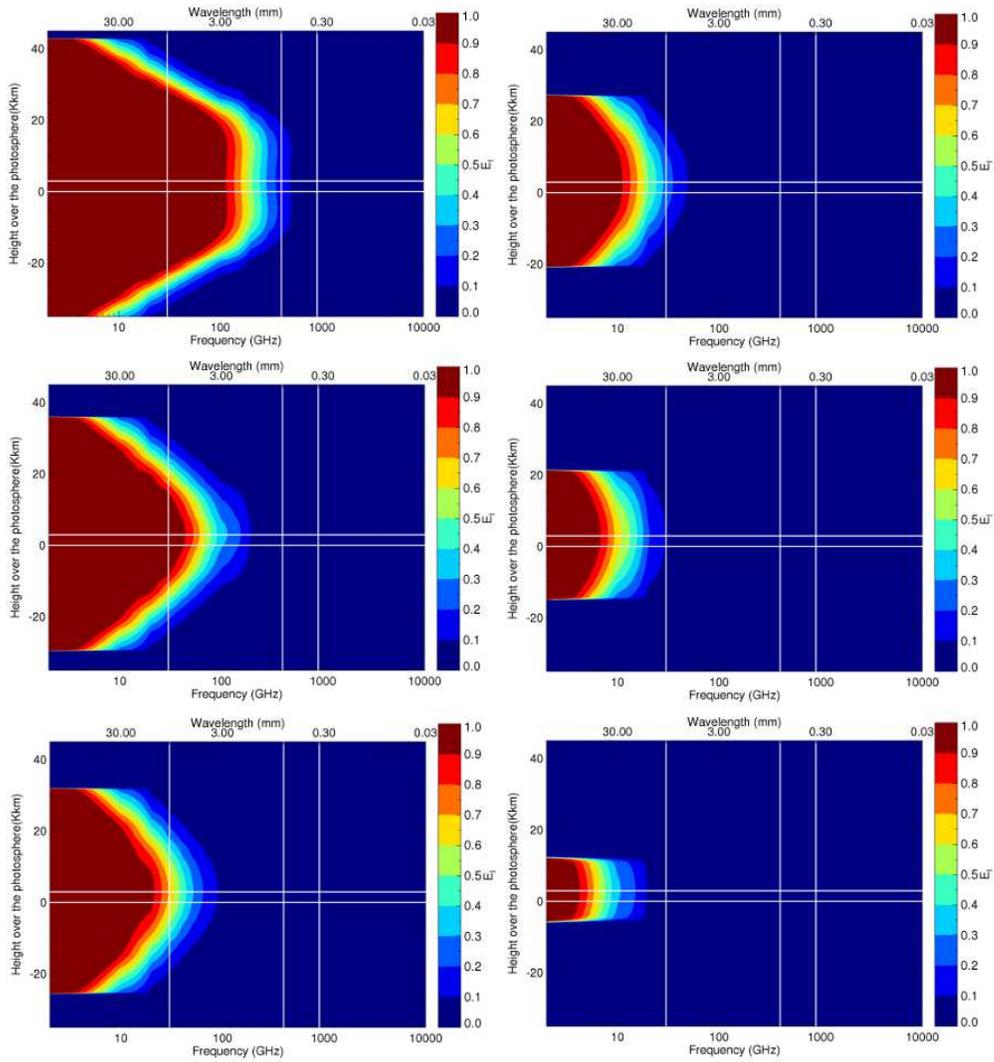}
\caption{Local emission efficiency from 1197 km (upper-left) to 2088 km
  (bottom-right) over the photosphere.}\label{figure2.ps}
\end{center}
\end{figure}

We compute the CSMC structure in 13 ray paths positions, with steps of 78
km between each position, from -51 km to 2088 km over the photosphere (We plot
the last 12 ray paths in the Figures \ref{figure1.ps} and \ref{figure2.ps}) .  

For each ray path we compute the local emission efficiency
for frequencies between 2 GHz and 10 THz.
In this plot the ``y'' axe is the ``z'' coordinate defined in
\cite{2010ApJS..188..437D}, i.e. is the absolute position in the
line between the center of the sun and the observer. The colors
are the local emission efficiency.

We found that the CSMC change their morphology very close to the
limb.
% (see sub-figure 1 and 2 of the Figure \ref{figure1.ps}). 
In the case of 51 km below the photosphere at limb, we found that the CSMC
morphology  is very similar to the CSMC at the center of the solar disk.
The main difference is the depth of the structure, in this case the
CSMC have more than 40,000 km in depth while that in the center of the solar
disk have only 2100 km.

The CSMC structure at 127 km over the limb changes dramatically (sub-figure
1 of the Figure \ref{figure1.ps}). 
Their morphology presents 3 regions of high emission and the depth
is around 80,000 km. We observe an asymmetry in the ``y'' axe. This
asymmetry is originated by the 2D geometry of the model.

The three region characteristic is preserved between 127 km and 662 km over
the photosphere at the limb (sub-figures 1-4 of the Figure \ref{figure1.ps}),
we observe that the outer regions of the CSMC gradually fall to the center,
while the central peak decrease in frequency.

Between the 662 and 840 km the central peak disapear and only observe
two regions of emission that gradually comes one peak around 1018 km.
Finally, between 1018 and 2087 km the single structure gradually down in
frequency and in height.

\section{Discussion and Conclusion}
The morphology of the CSMC at the limb shows three steps, the first one
between -51 and 127 km over the photosphere, where the structure of the
CSMC changes dramatically and the central peak of the structure is higher than
the another two peaks around them. The second step is between 306 and
1375 km where the combination of the fall of the central peak in frequency
and the fall of the two outer peaks to the center produces a constant total
emission, i.e. while we observe away from the photosphere, the emission of
at frequencies between 2 and 500 GHz remains almost constant. Finally
between 1553.08 and 2087.74 km the emission fall in frequency.

These three steps in the CSMC at the limb shows that the solar radii have
three region: at low frequencies the solar radii downs slowly from 2 GHz to
50 GHz, then a region between 50 GHz to 1000 GHz where the solar radii remains
almost constant and finally a third region where the solar radii decreases at
photosphere altitudes after the 1000 GHz.

We can conclude that the CSMC is a tool that help us in the study of the limb
brightening and the solar radii. We also show that the regions of emission at
the limb have several changes in opacity and could be useful to characterize
the micro structure at this altitudes.

%Figures
%/home/vdelaluz/seris/departurecoefficient/TvsNH.ps

%\section{}   %%% Top level section head (remove "%" symbol)
%\label{}
%\subsection{}   %%% Second level section head (remove "%" symbol)
%\label{}
%\subsubsection{}   %%% Lowest (unnumbered) level section head (remove "%" symbol)
%\label{}
%\altsubsubsection{} %%% Lowest numbered level section head (remove "%" symbol)
%\label{}

%\acknowledgments %%% Text of acknowledgments runs on after this command.

%%% THE BIBLIOGRAPHY
%%%
%%% CONSULT SECTION 3.5 OF "INSTRUCTIONS FOR AUTHORS" FOR HOW TO USE NATBIB
%%% IN COMBINATION WITH BIBTEX.
%%%

\bibliography{author} %%%THIS CALLS THE BIBLIOGRAPHICAL DATABASE FILE author.bib.

\end{document}